\documentclass[sigconf]{acmart}

\pdfoutput=1

\makeatletter
\def\mdseries@tt{m}
\makeatother

\usepackage{booktabs}
\usepackage{multirow}
\usepackage[nomain,acronym,section]{glossaries}
\usepackage{siunitx}
\usepackage[lofdepth,lotdepth]{subfig}

\usepackage{pgfplots}
\pgfplotsset{compat=1.14}

\usepackage{tikz}
\usetikzlibrary{arrows}
\usetikzlibrary{automata}
\usetikzlibrary{backgrounds}
\usetikzlibrary{calc}
\usetikzlibrary{fit}
\usetikzlibrary{matrix}
\usetikzlibrary{patterns}
\usetikzlibrary{positioning}
\usetikzlibrary{shadows}

\usepackage{todonotes}
\usepackage[roman]{parnotes}

\usepackage[xlatin,abbreviations]{foreign}

\ifdefined\ReviewMode
    \setcopyright{none}
\else
    \setcopyright{licensedothergov}
\fi

\acmConference{ACSAC 2017}{December 4--8, 2017}{Orlando, FL, USA}
\acmYear{2017}
\copyrightyear{2017}
\acmPrice{15.00}
\acmDOI{10.1145/3134600.3134622}
\acmISBN{978-1-4503-5345-8/17/12}

\glsdisablehyper
\newacronym{BIOS}{BIOS}{Basic Input/Output System}
\newacronym{UEFI}{UEFI}{Unified Extensible Firmware Interface}
\newacronym{DXE}{DXE}{Driver Execution Environment}
\newacronym{OS}{OS}{Operating System}
\newacronym{NAVIS}{NAVIS}{Network Adapter Verification and Integrity Solution}
\newacronym{ROP}{ROP}{Return-Oriented Programming}
\newacronym{JOP}{JOP}{Jump-Oriented Programming}
\newacronym{TCB}{TCB}{Trusted Computing Base}
\newacronym{SMM}{SMM}{System Management Mode}
\newacronym{SMRAM}{SMRAM}{System Management RAM}
\newacronym{SMRR}{SMRR}{System Management Range Register}
\newacronym{SMI}{SMI}{System Management Interrupt}
\newacronym{PCI}{PCI}{Peripheral Component Interconnect}
\newacronym{TPM}{TPM}{Trusted Platform Module}
\newacronym{DMA}{DMA}{Direct Memory Access}
\newacronym{IDT}{IDT}{Interrupt Descriptor Table}
\newacronym{DKOM}{DKOM}{Direct Kernel Object Manipulation}
\newacronym{KOH}{KOH}{Kernel Object Hooking}
\newacronym{DFI}{DFI}{Data-Flow Integrity}
\newacronym{CFI}{CFI}{Control-Flow Integrity}
\newacronym{CFG}{CFG}{Control-Flow Graph}
\newacronym{SBAP}{SBAP}{Software-Based Attestation for Peripherals}
\newacronym{ATRA}{ATRA}{Address Translation Redirection Attack}
\newacronym{UID}{UID}{User Identifier}
\newacronym{BIOSCNTL}{BIOS\_CNTL}{BIOS Control register}
\newacronym{BIOSWE}{BIOSWE}{BIOS Write Enable}
\newacronym{TOCTOU}{TOCTOU}{Time Of Check to Time Of Use}
\newacronym{IR}{IR}{Intermediate Representation}
\newacronym{IPC}{IPC}{Inter-Process Communication}
\newacronym{QPI}{QPI}{QuickPath Interconnect}
\newacronym{FPGA}{FPGA}{Field-Programmable Gate Array}
\newacronym{BITS}{BITS}{BIOS Test Suite}
\newacronym{LTO}{LTO}{Link-Time Optimization}
\newacronym{DOS}{DOS}{Denial Of Service}
\newacronym{FIFO}{FIFO}{First In First Out}
\newacronym{NIC}{NIC}{Network Interface Controller}
\newacronym{CRTM}{CRTM}{Core Root of Trust for Measurement}

\newcommand{\specialcell}[2][c]{\begin{tabular}[#1]{@{}c@{}}#2\end{tabular}}

\newenvironment{customlegend}[1][]{%
    \begingroup
    \csname pgfplots@init@cleared@structures\endcsname
    \pgfplotsset{#1}%
}{%
    \csname pgfplots@createlegend\endcsname
    \endgroup
}%

\def\addlegendimage{\csname pgfplots@addlegendimage\endcsname}

\pgfdeclarelayer{foreground}
\pgfsetlayers{background,main,foreground}
\tikzset{
    -|-/.style={
        to path={
            (\tikztostart) -| ($(\tikztostart)!#1!(\tikztotarget)$) |- (\tikztotarget)
            \tikztonodes
        }
    },
    -|-/.default=0.5,
    |-|/.style={
        to path={
            (\tikztostart) |- ($(\tikztostart)!#1!(\tikztotarget)$) -| (\tikztotarget)
            \tikztonodes
        }
    },
    |-|/.default=0.5,
}

\begin{document}
\title[Co-processor-based Behavior Monitoring: Detection of Attacks Against the SMM]{Co-processor-based Behavior Monitoring: Application to the Detection of Attacks Against the System Management Mode}

\author{Ronny Chevalier}
\affiliation{\institution{HP Labs}}
\email{ronny.chevalier@hp.com}

\author{Maugan Villatel}
\affiliation{\institution{HP Labs}}
\email{maugan.villatel@hp.com}

\author{David Plaquin}
\affiliation{\institution{HP Labs}}
\email{david.plaquin@hp.com}

\author{Guillaume Hiet}
\affiliation{\institution{CentraleSup\'elec}}
\email{guillaume.hiet@centralesupelec.fr}

\renewcommand{\shortauthors}{Chevalier et al.}

\begin{abstract}
Highly privileged software, such as firmware, is an attractive target for attackers. Thus,
BIOS vendors use cryptographic signatures to ensure firmware integrity at boot time.
Nevertheless, such protection does not prevent an attacker from exploiting vulnerabilities at runtime.
To detect such attacks, we propose an event-based behavior monitoring approach that relies
on an isolated co-processor.
We instrument the code executed on the main CPU to send information about its behavior to the monitor.
This information helps to resolve the semantic gap issue.
Our approach does not depend on a specific model of the behavior nor on a specific target.
We apply this approach to detect attacks targeting the
\acrfull{SMM}, a highly privileged x86 execution mode executing firmware code at runtime.
We model the behavior of SMM using invariants of its control-flow and relevant CPU registers (CR3 and SMBASE).
We instrument two open-source firmware implementations: EDK~II and
coreboot. We evaluate the ability of our approach to detect state-of-the-art attacks and its runtime execution overhead 
by simulating an x86 system coupled with an ARM Cortex A5 co-processor. The results show that
our solution detects intrusions from the state of the art, without any false positives, while remaining acceptable in 
terms of performance overhead in the context of the \acrshort{SMM} (\ie less than the \SI{150}{\micro\second} threshold defined by Intel).
\end{abstract}

%
%



\maketitle

\section{Introduction}
    \label{sec:introduction}
    Computers often relies on low-level software, like the kernel of an \acrfull{OS} or software embedded in the hardware, 
called firmware. Due to their early execution and their direct access to the hardware, these low-level components are 
highly privileged programs. Hence, any alteration to their expected behavior, malicious or not, can have dramatic 
consequences on the confidentiality, integrity or availability of the system.

Boot firmware, like the \acrfull{BIOS} or \acrfull{UEFI} compliant firmware, is in charge of testing and initializing 
hardware components before transferring the execution to an OS\@. In addition to boot firmware, the platform
initializes and executes runtime firmware code while the OS is running. On x86 systems, a highly privileged 
execution mode of the CPU, the \acrfull{SMM}~\cite{intelsysSMM}, executes runtime firmware code.

Any attacker that can change the original behavior of boot or runtime firmware, like skipping a verification step, can compromise the system. For this reason, tampering with the firmware is 
appealing for an attacker and sophisticated malware tries to infect it. Such malware is persistent, hard to detect, and does not depend on the OS installed on the platform~\cite{embleton2013smm,ars2013SMM,trend2013HT}.

Firmware code is stored on dedicated flash memory. On x86 systems, only runtime firmware code executed in SMM is 
allowed to modify the flash. It prevents a compromised OS from infecting the firmware.
During the boot
phase and before executing the \acrshort{OS}, the boot firmware loads some code in \acrfull{SMRAM}. This code
corresponds to privileged functions that will be executed in SMM\@. Then, the firmware locks the
\acrshort{SMRAM} and the flash (using hardware features) to prevent any modification from the OS\@.
Furthermore, recent firmware uses cryptographic signatures during the boot process~\cite{pi2016,ruan2014intel,hp2016ss}
and the update process~\cite{cooper2011bios} to ensure that only firmware signed by the vendor's key is updated and is 
executed. In
addition, measurements (cryptographic hash) of all the components and configurations of the boot process can be computed
and securely stored at boot time, to attest the integrity of the platform~\cite{trusted2007tpm}.

While cryptographic signatures and measurements 
provide code and data integrity at boot time, they do not prevent an attacker 
from exploiting a vulnerability in \acrshort{SMM} at runtime~\cite{wojtczuk2009attackingSMM,duflot2009getting,intel2015new}. Hence, we 
need ways to prevent vulnerabilities in \acrshort{SMM}, or at least to detect intrusions exploiting such 
vulnerabilities.

Our work focuses on designing an event-based monitor for detecting 
intrusions that modify the expected behavior of the \acrshort{SMM} code at runtime.
While monitoring the behavior of \acrshort{SMM} is our primary goal, ensuring the integrity of the monitor itself is critical to prevent an
attacker from evading detection. Thus, we isolate the monitor from the monitored component (\ie the target) by using a
co-processor.

A common issue affecting hardware-based approaches that rely on an isolated monitor is the semantic 
gap between the monitor and the target~\cite{2004copilot,lee2013ki,mguard}. Such semantic gap issue 
occurs when the monitor only has a partial view of the target state. For example, if the monitor
gets a snapshot of the physical memory without knowing virtual to physical mapping (\eg CR3 register 
value on x86) it cannot reconstruct accurately the memory layout of the target.
Our monitor addresses this issue by leveraging a communication channel that allows the target to send
any information required to bridge this 
semantic gap. We enforce the communication of information relevant to the detection method via an 
instrumentation phase.
In addition, we ensure that the attacker cannot forge messages without first 
being detected.

Our detection approach relies on a model of the expected behavior of the monitored component,
while any significant deviation from this behavior is flagged as illegal. We chose an anomaly-based approach as we aim to 
detect exploits of unknown vulnerabilities.

In summary, our approach consists in detecting malicious behavior of a target program executed on a main CPU\@. The
detection is implemented in a monitor executed on an isolated co-processor. We also instrument the target code to 
enforce the communication between the target and the monitor at runtime.

This approach can be applied to monitor various low level software, such as SMM or ARM TrustZone secure 
world~\cite{ARMTZ}, which have the following properties: expose primitives called infrequently by upper layers and
perform minimal computation per primitive.
Moreover, 
different detection approaches could be used.
While generic, such approach introduces multiple challenges (\eg the overhead involved by the communication, the
provenance of the messages, or the integrity of the code added by the instrumentation phase).
In this paper,
we focus on the detection of attacks targeting the SMM code as a use case and show how we tackled these challenges.
We enforce
\acrfull{CFI}~\cite{wang2010hypersafe,tice2014enforcing,abadi2005control,zhang2013control,grsecRAP,carlini2015control,
niu2014modular,burow2017control} and monitor the integrity of relevant CPU registers (CR3 and SMBASE)
to illustrate the feasibility of our approach.

Our contributions are the following:
\begin{itemize}
 \item We propose a new approach using an event-based monitor targeting low-level software (\autoref{sec:approach}).
 \item We study the applicability of our approach using CFI to detect attacks against SMM runtime firmware code 
 (\autoref{sec:implementation}).
 \item We develop a prototype implementing our approach.
 \item We evaluate our approach in terms of detection capability and performance overhead on real-world firmware widely 
 used in the industry (\autoref{sec:evaluation}).
\end{itemize}

This paper is structured as follows. First, in \autoref{sec:approach}, we provide an overview of our
generic approach. Then, in \autoref{sec:background}, we give a brief
background on \acrshort{CFI} and SMM\@. We detail the threat model associated with this
use case in \autoref{sec:threat_model}. We describe the design
and implementation of our prototype in \autoref{sec:implementation}.
In~\autoref{sec:evaluation}, we evaluate our approach. In
\autoref{sec:related_work}, we compare our approach with related work. Finally, we
conclude and propose some future work in~\autoref{sec:conclusion}.

\section{Approach Overview \& Requirements}
    \label{sec:approach}
    In this section, we describe the generic concepts and requirements of our event-based behavior monitoring approach.
As explained in~\autoref{sec:introduction}, such concepts could be used to monitor different targets and could rely on
different detection approaches. We detail in~\autoref{sec:implementation} one possible implementation of this generic
approach to detect runtime attacks on SMM code using CFI\@.

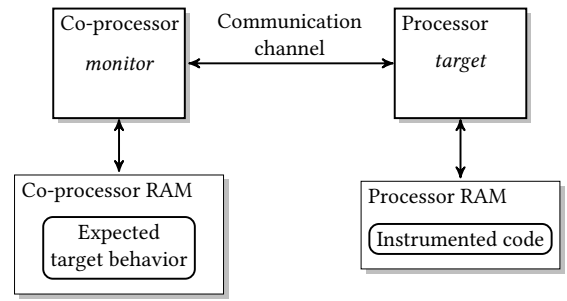
\begin{figure}[ht]
  \centering
  \resizebox{0.9\columnwidth}{!}{
  \tikzstyle{ram} = [draw, text centered,minimum width=7.5em, minimum height=4em, fill=white, drop shadow]
\tikzstyle{processor} = [draw, text centered,minimum width=6em, minimum height=5em,fill=white,drop shadow]
\begin{tikzpicture}[<->,>=stealth',shorten >=1pt,auto,thick]
    \node[processor,minimum width=6em] (coproc) {\textit{monitor}};
    \node[processor,right=3cm of coproc] (proc) {\textit{target}};
    \node[below right] at (coproc.north west) {Co-processor};
    \node[below right] at (proc.north west) {Processor};

    \node[draw,rounded corners,below=1.4cm of coproc,align=center] (policy) {Expected\\target behavior};

    \node[draw,rounded corners,below=1.5cm of proc] (code) {Instrumented code};

    \begin{scope}[on background layer]
        \node[ram, fit=(policy), inner sep=4mm,below=.8cm of coproc, minimum width=8.4em] (DRAM1) {};
        \node[below right,align=center] at (DRAM1.north west) {Co-processor RAM};
        \node[ram,fit=(code),yshift=2mm] (DRAM2) {};
        \node[below right,align=center] at (DRAM2.north west) {Processor RAM};
    \end{scope}

    \draw (coproc) -- (DRAM1);
    \draw (proc) -- (DRAM2);

    \draw (proc) -- node [above,align=center] {Communication\\channel} (coproc);
\end{tikzpicture}
  }
  \caption{High-level overview of the approach}
  \label{fig:approach_overview}
\end{figure}

Our approach, illustrated in~\autoref{fig:approach_overview},
relies on three key components, which we detail in the following subsections: a co-processor, a communication channel,
and an instrumentation step.
The co-processor isolates the monitor from the target. The target uses the communication channel to give more precise 
information about its behavior to the monitor. The instrumentation step 
enforces the communication.

\subsection{Co-processor}
The integrity of the monitor is crucial, because it is a trusted component that we rely on to detect
intrusions in our system. The monitor could also be used to start remediation strategies and restore the system to a 
safe state. If the attacker compromised our monitor, we could not trust the detection nor the remediation.

When the target and the monitor share the same resources (\eg CPU or memory), it gives the attacker 
a wide attack surface. Thus, it is necessary to isolate the monitor from the target. Modern CPUs provide hardware 
isolation features (\eg SMM or ARM TrustZone~\cite{ARMTZ}) reducing the attack surface. However, if one wants to monitor the code 
executed in such environment, the monitor itself cannot benefit from these isolation features.

In our approach, we use a co-processor to execute the monitor. Such co-processor
has its own execution environment and memory. Thus, the attacker cannot directly access this dedicated memory even if 
the target has been compromised. The attacker could only influence the behavior of the monitor via the
communication channel, which becomes the only remaining attack surface. 
The simplicity of such an interface, however, makes it harder to find vulnerabilities and to attack the monitor.
Such design reduces the attack surface.

In the following subsection, we discuss the requirements for our communication channel.

\subsection{Communication with the monitor}
\label{subsec:communication_monitor}
Being isolated from the target, the monitor cannot retrieve entirely the execution context of the target. Thus, there is 
a semantic gap between the current behavior of the target and what the monitor, executed on the co-processor, can infer 
about this behavior~\cite{jang2014atra,azab2010hypersentry}. For example, the monitor does not have sufficient 
information to infer the virtual to physical address mapping, nor the execution path taken at any point in time.

We introduce a communication channel between the monitor and the target. It allows the target to send messages to the
monitor.
Different types of information could be sent using this communication channel such as the content of a variable in 
memory, the content of a register, or the address of a variable. The nature of such information depends on the
detection approaches implemented on the monitor, providing flexibility in our approach.

The communication channel is the only remaining attack vector
against the monitor. Thus, how the monitor processes the messages and how the target
sends them are an important part of the security of the approach. To this end, we
require the following properties:

\begin{description}
 \item[(CC1) Message integrity] If a message is sent to the monitor, it cannot be removed or modified. Otherwise,
 an attacker could compromise the target and then modify or delete the messages before they are processed by the
 monitor to hide the intrusion.
 \item[(CC2) Chronological order] Messages are retrieved by the monitor in the order of
 their emission. Otherwise, an attacker could rearrange the order to evade the detection.
 \item[(CC3) Exclusive access] The instrumented code has exclusive access to the communication channel. Otherwise, an attacker could forge messages faking a legitimate behavior.
 \item[(CC4) Low latency] Sending a message should be fast (\eg sub-microsecond), because low-level components need to
minimize the time spent performing their task to avoid impacting higher-level components and the user experience.
\end{description}

\subsection{Instrumentation of the target}
\label{subsec:approach:instrumentation}
We enforce the communication from the target to the monitor by adding the communication code during an instrumentation 
step. This instrumentation step can be performed during the compilation or by rewriting the executable binary code.

Our approach relies on this enforcement, as should an attacker tamper with the 
instrumentation, the monitor would get inaccurate context of the behavior of the target making 
avoiding detection possible. Thus, the integrity of the instrumentation (\ie the communication code of the target) is
crucial. To this end, we require the following properties:
\begin{description}
    \item[(I1) Boot time integrity] The code and data at boot time are genuine and cannot be tampered with by the 
    attacker.
    \item[(I2) Runtime code integrity] The code cannot be modified by the attacker at runtime.
\end{description}

\section{Background}
    \label{sec:background}
    In this section, we provide an overview on control-flow hijacking and CFI\@. Then, we give some background regarding the
SMM\@.

\subsection{Control Flow Integrity (CFI)}
\label{subsec:cfi}
Widely used defense mechanisms such as non-executable data and non-writable executable code impede
attackers in their ability to exploit low-level vulnerabilities. Nevertheless, if an attacker managed to modify an
instruction pointer due to a vulnerability, then program execution would be compromised. For example, in an x86
architecture, programs store the return address of function calls on the stack. An attacker could exploit a
buffer overflow to overwrite the return address with an arbitrary one that redirect the execution flow. Code-reuse 
attacks, such as \acrfull{ROP}~\cite{roemer2012return} or
\acrfull{JOP}~\cite{bletsch2011jump,checkoway2010return}, use indirect branch instructions (\ie indirect call to a
function, return from a function and indirect jump) to chain together short instruction sequences of the existing code
to perform arbitrary computations.

The enforcement of a policy over the control-flow can prevent such attack. This defense mechanism, called 
\acrfull{CFI}, enforces integrity properties for each indirect branch where the control-flow 
transfer is determined at runtime. It ensures that a given execution of a program follows only paths defined by a 
\acrfull{CFG}. This graph represents all the legitimate paths that the program can follow. The \acrshort{CFG} needs to 
be defined ahead of time and it can be computed via source code analysis~\cite{abadi2005control}, binary
analysis~\cite{zhang2013control}, or execution profiling~\cite{yubin2012CFIMon}.

A typical way to enforce \acrshort{CFI} is by instrumenting the code, \eg during the compilation phase. This 
inlined-based approach adds runtime checks before each indirect 
branch~\cite{abadi2005control,wang2010hypersafe,tice2014enforcing}. If the address
is not within a finite set of allowed targets, the program stops.

A fined-grained CFI combines a shadow call stack (\ie an independent protected stack that only stores return
addresses) and a precise \acrshort{CFG} (\ie a \acrshort{CFG} with a small approximation regarding indirect
branches) to enforce \acrshort{CFI} on all indirect control transfers.

Some 
implementations~\cite{zhang2013control,fratric2012ropguard} sacrifice security over 
performance by building a less precise CFI\@. They either focus on protecting the backward-edge on the
\acrshort{CFG} (\eg with a shadow call stack) or on protecting the forward-edge
(\eg indirect calls). \citet{davi2014stitching} demonstrated that such implementations, called
coarse-grained CFI, fail to protect against control-flow hijacking. \citet{carlini2015control} also raised 
awareness on this issue by consolidating the argument that
\textit{without} stack integrity (\ie without using a shadow call stack), \acrshort{CFI} is insecure.

Our solution uses a type-based CFI inspired by the work of 
\citet{niu2014modular} and \citet{tice2014enforcing}. We implement a shadow call stack and verify 
that each indirect call branches to a function with an expected type signature known at compile time (more details
in~\autoref{sec:implementation}).

\subsection{System Management Mode (SMM)}
\label{subsec:smm}
\acrshort{SMM}~\cite{intelsysSMM} is a highly privileged execution mode of x86 processors. It provides the ability 
to implement 
\acrshort{OS}-independent functions (\eg advanced power management, secure firmware update, or 
configuration of UEFI secure boot variables)~\cite{intelsysSMM,yaotour}. The 
particularity of the \acrshort{SMM} is that it provides a separate execution environment, invisible to the 
\acrshort{OS}. The code and data used in SMM are stored in a hardware-protected memory region only accessible in 
\acrshort{SMM}, called SMRAM\@. \acrshort{SMM} is entered by generating a \acrfull{SMI}, which is 
a hardware interrupt. Software can also make the hardware trigger an SMI\@.

Access to the SMRAM depends on the configuration of the memory controller, done by the firmware during the boot.
Once all the necessary code and data have been loaded in SMRAM, the firmware locks the memory region so that it
can only be accessed by code running in \acrshort{SMM}, thus preventing an OS from accessing it. In 
addition, only the 
code executed in \acrshort{SMM} can modify the firmware stored into flash to 
prevent malware, executing with kernel privileges, from overwriting 
the firmware and becoming persistent.

The particularity of an \acrshort{SMI} is that it makes all the CPU cores enter \acrshort{SMM}. It is 
non-maskable and non-reentrant. Hence, this interrupt must be processed as fast 
as possible, since the \acrshort{OS} is paused during the handling of an SMI\@.

Despite hardware-based protection of the \acrshort{SMRAM}, several
attacks~\cite{wojtczuk2009attackingSMM,duflot2009getting,calloutofSMM2009,wojtczuk2009attacking,thinkpwn,intel2015new,
bulygin2017attacking,nucSMMVuln,sogetiSMMVuln} were publicly disclosed.
These attacks are proof-of-concepts that
attackers could use to perform arbitrary code execution in \acrshort{SMM}, once the SMRAM has been locked.

\paragraph{Cache poisoning}
Two research teams~\cite{wojtczuk2009attackingSMM,duflot2009getting} independently discovered
cache poisoning attacks in SMM\@. Since the cache is shared
between all the execution modes of the CPU, the attack consists in marking the \acrshort{SMRAM} region to be cacheable 
with a write-back strategy. Then, the attacker stores in the cache malicious instructions. After that, once an SMI is 
triggered, the processor fetches the instructions from 
the cache. Thus, the processor executes the malicious instructions of the attacker
instead of the legitimate code stored in SMRAM\@. The solution is to separate the cache between non-SMM and SMM
executions. This vulnerability has been fixed by adding a special-purpose register. Such register can only be 
modified in \acrshort{SMM} and decides the cache strategy of the SMRAM\@.

\paragraph{Insecure call}
Multiple firmware implementations~\cite{calloutofSMM2009} used call instructions to jump to code segments outside of the
SMRAM\@. An attacker with kernel-level privileges can easily modify this code. These vulnerabilities have been
fixed by forbidding the processor to execute instructions located outside of the \acrshort{SMRAM} while in SMM\@.

Other vulnerabilities due to indirect calls~\cite{thinkpwn,wojtczuk2009attacking} allow attackers to perform
code-reuse attacks against the \acrshort{SMM} code. Such attacks are usually prevented by patching these 
vulnerabilities. Our approach can detect code-reuse attacks in general without requiring patching.

\paragraph{Unchecked data}
Some \acrshort{SMI} handlers rely on data provided by the \acrshort{OS} (\ie controlled by the attacker). If they do
not sanitize such data, the attacker can influence the behavior of the SMM\@.

For example, pointer vulnerabilities in an \acrshort{SMI} handler can
lead to arbitrary write into SMRAM~\cite{nucSMMVuln,sogetiSMMVuln,intel2015new}. It can occur because the
\acrshort{SMI} handler writes data into a buffer located at an address controlled by the attacker. For example,
such address can be provided thanks to a register that could have been modified by the attacker.
\citet{bulygin2017attacking} also demonstrated a similar attack by modifying the Base Address Registers (BAR) used to 
communicate with PCI devices.

It is the responsibility of \acrshort{SMI} handlers to verify that the data given or controlled by the \acrshort{OS} is 
valid. For example, they should check that the address of the communication buffer is not pointing into the 
\acrshort{SMRAM}, and that the BARs point to valid addresses (\ie not in RAM or SMRAM).

\section{Threat model and assumptions}
    \label{sec:threat_model}
    As explained previously, the \acrshort{SMM} is the last bastion of firmware security. It is the only mode that
allows write access to the flash storage of the firmware, and its execution is invisible to the \acrshort{OS}, thus 
a perfect place to hide malware~\cite{embleton2013smm}.
In addition, it allows the attacker to perform actions
that cannot be realized with kernel privileges. For example, if the attacker wants to remain persistent or modify security 
configurations (\eg disable secure boot).

Every time a vulnerability related to the \acrshort{SMM} has been reported it has
been patched. Firmware, however, is not updated frequently~\cite{kovah2015many}. Moreover, in practice, vendors 
typically use third-party code to build their firmware making code review and
vulnerability management more difficult.

Hence, we assume that the attacker will find a vulnerability, but exploitation of such vulnerability 
implies a deviation from the expected behavior of the SMM code. Thus, our approach focuses
on monitoring its behavior. Such anomaly-based approach is not limited to the detection of well-known attacks, but can 
also detect the exploitation of unreported (zero-day) vulnerabilities.

We assume that the code during the boot process is legitimate and that no attack is performed during
that phase until the \acrshort{SMRAM} is locked.
Such an assumption is reasonable with the use of existing security mechanisms for recent firmware such as:
\begin{itemize}
    \item An immutable hardware root of trust to verify that the boot firmware has a valid cryptographic signature from the vendor before its execution~\cite{ruan2014intel,hp2016ss},
    \item Cryptographic signatures during the update process~\cite{cooper2011bios},
    \item A \acrfull{TPM} chip to measure all the components of the boot process at boot
    time~\cite{trusted2007tpm}.
\end{itemize}

These mechanisms provide us with code and data integrity at boot time (I1, a requirement stated 
in~\autoref{subsec:approach:instrumentation}).
In addition, since recent firmware use page tables~\cite{phoenixPageTable} in SMM
we can enable write protection~\cite{yao2015white,yao2015whiteMemory} and assume code integrity at runtime (I2).

Another key assumption is that the attacker cannot send messages in lieu of SMM without being detected.
First, by design, messages cannot be sent by other components than the CPU
and among the messages sent by the CPU only those sent in SMM are processed by the monitor (see~\autoref{subsubsec:implementation:channel:fifo}).
Second, we assume that there is no vulnerability in SMM code that can be exploited by an attacker
to forge messages without altering the control flow.
Since any attempt
to alter the control flow results in the emission of a message describing an invalid control flow
(see~\autoref{subsubsec:implem:instru:cfi}),
the attacker cannot forge messages without first being detected.

Finally, we do not consider an attacker trying to impede the availability of the system (denial of service) by flooding
the communication channel. The attacker already has sufficiently high privileges to perform a denial of service
(\eg shutdown the machine).
We model such an attacker with the following capabilities:

\begin{itemize}
  \item Complete control over the \acrshort{OS} or the hypervisor, meaning that the attacker already found 
vulnerabilities that elevate its privileges to kernel-level or hypervisor-level,
  \item Complete control over the memory, except the \acrshort{SMRAM}, which is protected,
  \item Cannot exploit hardware vulnerabilities (\eg cache poisoning attacks~\cite{wojtczuk2009attackingSMM,duflot2009getting} or bypassing SMRAM protection),
  \item Can trigger as many SMIs as necessary,
  \item Can exploit a memory corruption issue in an \acrshort{SMI} handler.
\end{itemize}

This threat model is close to those used in the different attacks described in~\autoref{subsec:smm} 
(except for the cache poisoning attack).

\section{SMM Behavior Monitoring}
    \label{sec:implementation}
\begin{figure}[ht]
    \centering
    \resizebox{\columnwidth}{!}{
        \tikzstyle{processor} = [draw, text centered]
\tikzstyle{queue} = [rectangle split, rectangle split parts=3, draw]
\begin{tikzpicture}[->,>=stealth',shorten >=1pt,auto,node distance=2cm,thick]
    \node [draw, align=center] (indCall) {Indirect calls\\handling};
    \node [draw,right=1em of indCall, align=center] (shadow) {Shadow\\call stack};
    \node [draw,right=1em of shadow, align=center] (other) {CPU registers\\integrity};

    \node [draw,below of=shadow] (fetchMsg) {Fetch message};

    \node [draw,above=6mm of shadow,rounded corners] (invalid) {Invalid?};

    \node [draw, above=2cm of invalid] (firmware) {Firmware};

    \begin{scope}[on background layer]
        \node [processor,fit=(fetchMsg) (shadow) (indCall) (other) (invalid), inner sep=3mm, drop shadow, fill=white] (monitor) {};
        \node [above right] at (monitor.south west) {Monitor};

        \node [processor,fit=(firmware), inner sep=3mm, yshift=2mm, fill=white, drop shadow] (target) {};
    \end{scope}
    \node [queue,above=1mm of monitor, xshift=3.5cm] (queue) {$\cdots$\nodepart{two}packet\nodepart{three}packet};
    \node [left,align=center] at (queue.west) {Memory\\mapped\\device};

    \node [below right] at (target.north west) {Target};

    \node [draw, left=4cm of firmware,align=center,yshift=-0.7cm] (compilation) {LLVM\\compilation};
    \node [draw, above of=compilation,node distance=1.1cm] (sourcecode) {Source code};

    \draw (fetchMsg) -- (shadow);
    \draw (fetchMsg) to[|-|] node [below] {4. Dispatch} (indCall);
    \draw (fetchMsg) to[|-|] (other);

    \draw[dashed] (sourcecode) -- (compilation);
    \draw[dashed] (compilation) -| node[xshift=-2.5cm,above,align=center] {1. Instrumented code} (firmware.south west);
    \draw[dashed] (compilation) |- node[right,yshift=1cm,align=center] {1. Indirect calls\\mapping} (indCall);

    \draw (firmware) -| node[above left] {2. Push packet} (queue);

    \draw (queue) |- node [above left, align=center, xshift=-5] {3. Pop\\packet} (fetchMsg);
    
    \draw (shadow) to[|-|] (invalid);
    \draw (indCall) |- node [above] {5. Detect} (invalid);
    \draw (other) |- (invalid);

    \draw (invalid) -- node [left, align=center] {6. Remediation action} (target);

    \begin{customlegend}[legend cell align=left,
        legend style={at={(-1.7,-1.7)}},
        legend entries={
        \footnotesize Compile time,
        \footnotesize Runtime
        }]
        \addlegendimage{line legend,dashed}
        \addlegendimage{line legend}
    \end{customlegend}
\end{tikzpicture}
    }
    \caption{High-level overview of the implementation}
    \label{fig:architecture_overview}
\end{figure}
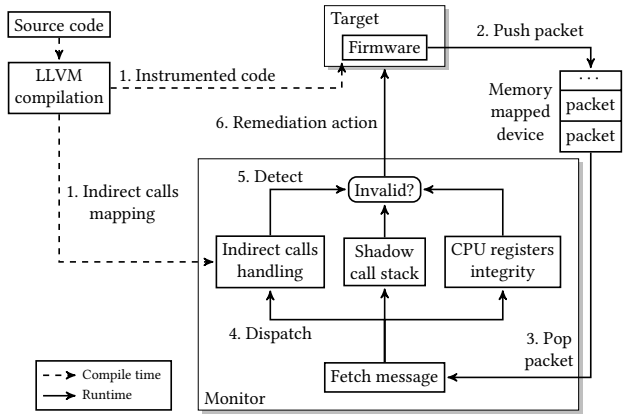

    We apply our generic approach to monitor the behavior of the SMM code using CFI and by
ensuring the integrity of relevant x86 CPU registers. The design of
our solution is illustrated in~\autoref{fig:architecture_overview}. In this
figure, straight arrows represent the steps taken during runtime and dashed arrows the steps taken 
during the instrumentation phase (compilation time). We describe our implementation in more details
in the following subsections.

\subsection{Detection method}

\subsubsection{CFI}
\label{subsubsec:implementation:detection:cfi}
We enforce a \acrshort{CFI} policy, because it is suited to detect attacks on low-level vulnerabilities
that often appears in code written in C\@.
Our monitor, executed on the co-processor, verifies that the control-flow information
sent by the target is valid.

The monitor implements a type-based CFI inspired by the work of 
\citet{niu2014modular} and \citet{tice2014enforcing}. It ensures that the address used in an
indirect call matches the address of a function having an expected type signature known at compile time.
For example, the call site \verb|s->func(s, 1, "abc")| is an indirect call where \verb|func| has
\verb|int (*func)(struct foo*, int, char *)| as a type signature. Thus, the monitor ensures that the address
of \verb|func| used at that call site always points to a function having the same signature.
In addition, the monitor implements a shadow call stack to ensure the integrity
of return addresses on the stack.

A type-based CFI over-approximates the set of expected pointers with all functions with the 
same type signature.
In practice, type-based CFI gives small equivalence classes~\cite{burow2017control} where one
equivalence class contains all the possible targets for one call site.
An alternative could be to use a points-to analysis such as the work from \citet{DSA}.
This type of analysis can sometimes give precise results (\ie the complete set of pointers).
However, in practice, as shown by~\citet{evans2015control},
such analysis often fails to give the accurate set of pointers resulting in large equivalent classes
such as all the available functions in the program.

Finally, our approach isolates the detection logic, the model of the behavior, and the data structures 
(\eg shadow call stack and indirect call mappings) with the use of an isolated co-processor. It prevents
attackers from tampering with it. Thus, we provide a more robust CFI using external monitoring in comparison to 
inlined-based CFI~\cite{abadi2005control}.

\subsubsection{CPU registers integrity}
In addition to a CFI policy, the monitor ensures the integrity of relevant x86 CPU registers in SMM\@.
It stores expected values in its memory at boot time and verifies the values 
sent by the target at runtime.

When entering \acrshort{SMM}, the main CPU stores its context in the save state area, and restores it when 
exiting~\cite{intelsysSMM}. The location of the \acrshort{SMRAM}, called the SMBASE, is saved in the save state area. 
The processor uses the SMBASE every time an \acrshort{SMI} is triggered to jump to the \acrshort{SMM} entry point. 
Hence, it is possible for an \acrshort{SMI} handler to modify the SMBASE in the save state area, and the next time an 
\acrshort{SMI} is triggered, the processor will use the new SMBASE\@. Such behavior is genuine at boot time to relocate
the \acrshort{SMRAM} to another location in RAM\@. At runtime, however, there is no valid reason to do this. If an
attacker manages to change the SMBASE, it results in arbitrary code execution when the next SMI is triggered. Therefore,
the monitor ensures that the SMBASE value does not change between SMIs at runtime.

In addition, the monitor ensures the integrity of MMU-related registers, like CR3 (\ie an x86 register holding
the physical address of the page directory).
Such register is an interesting target for attackers~\cite{jang2014atra}. 
Thus, protecting its integrity is needed since recent firmware enable
protected mode and use page tables~\cite{phoenixPageTable,yao2015white,yao2015whiteMemory}.
These registers are reset at the beginning of each SMI with a
value stored in memory. Such value is not supposed to change at runtime. If an attacker succeeds in
modifying this value stored in memory, then the corresponding register is under the control of the 
attacker at the beginning of the next SMI\@.

\subsection{Co-processor}
\label{subsec:implem_coproc}
We take inspiration from the AMD Secure Processor, also known as the Platform Security
Processor (PSP)~\cite{amdpsp}, and the Apple Secure Enclave Processor (SEP)~\cite{mandtdemystifying}.
Both are used as a security processor to perform sensitive tasks and handle sensitive data (\eg cryptographic keys).
In those solutions, the main CPU cannot directly access the memory of the co-processor. It only asks
the co-processor to perform security-sensitive tasks via a communication channel.

The PSP is an ARM Cortex A5 and the SEP is an ARM Cortex A7. Such processors are similar, they are both 32 bit
ARMv7 with in-order execution and 8-stage pipeline. The main difference is that the A5 is
single-issue and the A7 is partial dual-issue.

In our implementation, we chose a similar design and we use an ARM Cortex A5 co-processor to execute 
our monitor. It gives us the isolation needed and enough processing power to process the messages for our use case.

We implemented our monitor with approximately 1300 lines of Rust~\cite{rust}, a safe system programming 
language.

\subsection{Communication channel}
\label{subsec:implem_channel}
In this subsection, we look at how existing co-processors communicate with the main CPU and explain why they do not
fit our requirements. Then, we describe how we design our communication mechanism to fulfill the properties we defined
in~\autoref{subsec:communication_monitor}.

\subsubsection{Existing mechanisms}
\label{subsubsec:implementation:channel:existing}
A major characteristic of the communication channel is its performance, especially its latency,
as each message sent impacts the overall latency of SMI handlers.

The Intel \acrfull{BITS} defined the acceptable latency
of an \acrshort{SMI} to \SI{150}{\micro\second}~\cite{intelBITS150}. 
\citet{delgado2013performance} showed that, if the latency exceeds this threshold, it causes a 
degradation of performance (I/O throughput or CPU time) or user experience (\eg severe drop in frame rates in game 
engines).

Both the PSP and the SEP use mailbox communication channels to send and receive messages 
with the main CPU~\cite{amd2016bkdg,mandtdemystifying}.
Mailboxes work as follows. One processor writes to a mailbox register, which triggers
an interrupt in a second CPU\@. Upon receiving the interrupt, the second CPU
executes code that fetches the value in the mailbox, processes the message, and then writes a
response.

We could use such a mechanism to fulfill our security properties (CC1 and CC2) by making the SMM code 
wait until the co-processor acknowledged the message.
\citet{shelton2013popcorn} studied the latency of mailboxes on Linux and
measured on average a 7500 cycles latency. For example, with a \SI{2}{\giga\hertz} clock this gives 
\SI{3.75}{\micro\second} per message. Thus, not fulfilling the low-latency requirement (CC4).

Since the mechanism used by existing co-processors, like the PSP or the SEP, does not allow low
latency communication while fulfilling our security requirements, we designed a specific hardware component to that end.

\subsubsection{Restricted FIFO}
\label{subsubsec:implementation:channel:fifo}
We designed a restricted \acrfull{FIFO} queue between the main CPU and the co-processor.
This \acrshort{FIFO} is implemented as an additional hardware component connected to the main CPU and
the co-processor, because we want to re-use existing processors without modifying them.

The goal of the FIFO is to store the messages sent by the target awaiting to be processed by the
co-processor. The \acrshort{FIFO} only allows the main CPU to push messages and the co-processor to pop
them. The FIFO receives messages fragmented in packets. Only our FIFO handles the storage of the
messages, the attacker does not have access to its memory, thus it cannot violate the integrity of
the messages. We consider single-threaded access to the \acrshort{FIFO}, since only one core
handles the SMI, while other cores must wait~\cite{intelsysSMM}.\footnote{At the beginning of each SMI,
there is a synchronization code ensuring that only one core executes in SMM\@. This implies that we do not
instrument the code responsible for the synchronization between the cores. Such code does not 
interact with any attacker-controlled data and cannot be influenced by the attacker, hence we trust 
it.}

We are using a co-processor with less processing power than the main CPU and the monitor usually 
processes messages at a lower rate than their production. Thus, the FIFO could overflow. Such a
case would happen if the monitored component would be continuously executing, which is not the case
with SMM code. Most of the time the main CPU will execute code in kernel land or
userland, which are not monitored and hence do not send any message. An SMI, on the other hand, will
create a burst of messages when triggered. Hence, the only case where
the FIFO could overflow is if an attacker deliberately triggered SMIs at very high rate,
which would be detected as an attack.

We use a fast interconnect between the main CPU executing the monitored component and the FIFO\@.
The precise interconnect depends on the CPU manufacturer. In the x86 world two major interconnects exist:
\acrfull{QPI}~\cite{intel2009QPI} from Intel and HyperTransport~\cite{holden2008hypertransport} from AMD\@.

These interconnects are used for inter-core or inter-processor communication and are specifically
designed for low latency. For example, CPU manufacturers are using them to maintain cache coherency. Furthermore, they
have been leveraged to perform CPU-to-device communication~\cite{litz2008velo,litz2010tccluster,froning2013achieving}.
The co-processor could be connected to the FIFO using these interconnects (using glue logic) or an interconnect
with similar performance (\eg AMBA~\cite{ARMAMBA}).

Our monitored component has a mapping between a physical address and the hardware
component (\ie the \acrshort{FIFO}) allowing it to send packets via the interconnect. Routing tables 
are used by interconnects. Such routing tables are configured via a software interface (with kernel 
privileges) to decide where the packets are sent. Thus, as explained
by \citet{song2014security}, it would be possible for an attacker to modify the routing tables to prevent the 
delivery of the messages to the FIFO\@. Such attack would be the premise of an
attack against a vulnerable SMI handler. Therefore, at the beginning of each \acrshort{SMI}, we 
enforce the mapping by overwriting the routing table in the SMM code to prevent such an attack.

In addition, the FIFO filters the messages by checking the 
SMIACT\# signal of the CPU specifying whether the main CPU is in SMM or not~\cite{holden2008hypertransport,intelsysSMM}.
Hence, the monitor only processes messages sent in SMM and prevents an attacker from sending messages when the target
is not executing (\eg an attacker sending messages in kernel mode).

To summarize, this design fulfills the message integrity property (CC1), since
the target can only push messages to the restricted FIFO\@. Moreover, if the queue is full it does
not wrap over and the target enforces the routing table mapping.
It fulfills the chronological order property (CC2), because it is a
FIFO and there is no concurrent access to it while in SMM\@.
In addition, it fulfills the exclusive access property (CC3), 
since we filter messages to ensure they only come from the SMM, the integrity of the instrumentation code
is ensured with the use of page tables with write-protection enabled, and the attacker cannot forge messages
without first being detected.
Finally, we fulfill the last property (CC4) by 
using a low latency interconnect between the main CPU and the FIFO\@.

\subsection{Instrumentation}
\label{subsec:implem:instru}
The instrumentation step of our implementation that modifies the SMM code is twofold: (1) an instrumentation 
to send CFI related information; and (2) an instrumentation to send information regarding x86 specific variables. 

As previously stated, the goal of the instrumentation is to \emph{send} information to the monitor. 
In comparison 
to other approaches where they inline some verifications in the instrumented code, we 
only use \verb~mov~ instructions to send packets to our FIFO\@.

\subsubsection{CFI}
\label{subsubsec:implem:instru:cfi}
We rely on LLVM 3.9~\cite{LLVM}, a compilation
framework widely used in the industry and the research community, to instrument the SMM code.
We implement two LLVM passes with approximately 600 lines of C++ code.

The first pass enforces the backward-edge \acrshort{CFI} (\ie a shadow call stack). It instruments the SMM code
to send one message at the prologue and epilogue of each function. Such message contains the return 
address stored on the stack.

The second pass enforces forward-edge \acrshort{CFI} (\ie indirect calls always branch to valid targets). 
For each indirect call site, we assign a unique identifier (CSID), we create a
mapping between their CSID and the type signature of the function called, and we add this type into a set of types
called indirectly (SIND). We
instrument each indirect call site to send the CSID and the branch target address to the monitor before executing
the indirect call.

Then, for each function whose type signature is in SIND, we add a mapping between the function offset in memory and its 
type. This mapping gives us all the functions that could be called indirectly with their type signature and offset 
in memory.

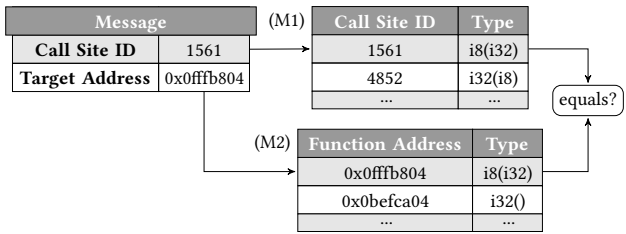
\begin{figure}[ht]
    \centering
    \resizebox{\columnwidth}{!}{
        \tikzset{
    table/.style={
        matrix of nodes,
        row sep=-\pgflinewidth,
        column sep=-\pgflinewidth,
        nodes={
            rectangle,
            draw=black,
            align=center
        },
        text depth=0.1ex,
        text height=1.8ex,
        nodes in empty cells,
        every even row/.style={
            nodes={fill=gray!20}
        },
        column 1/.style={
            nodes={text width=7em}
        },
        row 1/.style={
            nodes={
                fill=black!40,
                text=white,
                font=\bfseries
            }
        }
    }
}
\begin{tikzpicture}[->,>=stealth',shorten >=1pt,auto]
    \matrix (msg) [table,text width=4.1em, column 1/.style={nodes={text width=7em,font=\bfseries}}]
    {
        Message & \\
        Call Site ID  & 1561 \\
        Target Address & 0x0fffb804 \\
    };
    \node[fit=(msg-1-1)(msg-1-2),rectangle,draw=black,fill=black!40,
    text=white,font=\bfseries,text depth=0.1ex,text height=1.8ex,align=center]{Message};

    \matrix (uids) [table, yshift=-0.141cm, text width=3em, right=0.8cm of msg,row 4/.style={text height=0.5ex}]
    {
        Call Site ID   & Type \\
        1561  & i8(i32) \\
        4852  & i32(i8) \\
        ...   & ... \\
    };

    \matrix (addrs) [table, yshift=0.9cm, text width=3em,row 4/.style={text height=0.5ex,},column 1/.style={nodes={text width=8.5em}},below=of uids]
    {
        Function Address    & Type \\
        0x0fffb804 & i8(i32) \\
        0x0befca04 & i32() \\
        ...        & ... \\
    };

    \node[draw,rounded corners, yshift=0.8cm, below right=0.4cm of uids] (equals) {equals?};

    \draw[->] (uids-2-2.east) -| (equals);
    \draw[->] (addrs-2-2.east) -| (equals.south);

    \draw[->] (msg-2-2.east) -- (uids-2-1.west);
    \draw[->] (msg-3-2.south) |- (addrs-2-1.west);
    
    \node [left] at (addrs-1-1.west) {(M2)};
    \node [left] at (uids-1-1.west) {(M1)};
\end{tikzpicture}
    }
    \caption{Mappings used to verify indirect calls messages}
    \label{fig:mappings_monitor}
\end{figure}

At the end of the build process, we provide two pieces of information: (1) a mapping between a CSID and a
type; and (2) a mapping between an \emph{offset} and the type of the function at that location.
However, such information is not enough for the monitor to have the mapping at runtime. It only has the functions
offset and not their final address in memory, hence the monitor needs the base address of the code used for the SMM\@.
We provide this information to the monitor by instrumenting the firmware
code to send the address during the initialization phase (before the \acrshort{SMRAM} is locked).
This way, at boot time, the monitor computes the final mapping by adding the offset to the corresponding base address.

Finally, as illustrated in~\autoref{fig:mappings_monitor}, the monitor can compute two
mappings: (M1) a mapping between a CSID and its expected type; and (M2) a mapping between the address
of a function and its type. Thanks to this information, the monitor can verify that the target 
address received in a message has the expected type according to the call site ID from the same 
message. The attacker can control the target address, but not the call site ID\@.

\subsubsection{CPU registers integrity}
\label{subsubsec:implem:instru:invariants}
We also instrument the SMM code to send some values related to x86 CPU registers.
These values, such as SMBASE or the saved value of CR3, could be modified by an attacker to take control of the SMM or 
evade detection.

First, we add some code executed at boot time to send the current values to the monitor. Since
there is no legitimate modification of these values at runtime, the monitor registers them.
Secondly, we add some code executed at runtime to send the values at the end of each SMI\@.

\section{Evaluation}
    \label{sec:evaluation}
    We evaluated our approach on two real-world implementations of code running in 
SMM\@. We first conducted a security evaluation of our approach using QEMU, as described 
in~\autoref{subsec:evaluation:security}. Then, we used the gem5 simulator to evaluate the runtime overhead of our 
approach, as detailed in~\autoref{subsec:evaluation:performance}.

We used a simulation-based prototype in order to have enough flexibility in exploring the hardware architecture, in a 
manner that would have been difficult to achieve using real hardware, such as FPGA-based solutions.\footnote{
At the time of writing, to the best of our knowledge, there is no off-the-shelf FPGA-based solutions with direct access 
to HyperTransport or Intel QPI commercially available.} A simulation allows us to simulate an interconnect and
to simulate the delay it takes for the main CPU to send one packet to the restricted FIFO\@.

\subsection{Experimental setup}
We used EDK~II~\cite{edk2} and coreboot~\cite{coreboot},
two real-world implementations of code running in \acrshort{SMM}.
EDK~II is an open source \acrshort{UEFI} compliant firmware used as the foundation for most vendor-based firmware. 
Coreboot is an open source firmware performing hardware initialization before executing a payload (\eg 
legacy BIOS or UEFI compliant firmware).
We built this firmware using our LLVM toolchain and we only instrumented the SMM related code.

\subsubsection{Simulator and emulator}
We both used a simulator and an emulator to validate our approach. The main goal of emulators is to be 
as feature-compatible as possible. However, they are not cycle-accurate and does not try to model accurately the 
performance of x86 or ARM platforms. Simulators, on the other hand, try to model accurately the performance of the
platforms they simulate, but often do not implement all their features (\eg no possibility to lock the
\acrshort{SMRAM}). Therefore, we use emulators to have all the SMM features, which is mandatory for security
evaluation, and simulators to model accurately the performance of our implementation.

For the security evaluation, we used the QEMU 2.5.1~\cite{qemu} emulator.
We modified QEMU to emulate our communication channel.

We used the gem5~\cite{binkert2011gem5} cycle-accurate simulator to estimate the performance impact both on the main CPU 
by modeling an x86 system, and on the co-processor by modeling an ARM Cortex A5.
\citet{butko2016full} evaluated that gem5 gave a performance prediction with a 20\% error on average.

We modified gem5 to simulate our \acrshort{FIFO} communication channel. It allowed us to
specify the delay (in nanoseconds) it takes to send or receive information from it.
We give the parameters used for gem5 in~\autoref{appendix:gem5}.

\subsubsection{Simulated communication channel delay}
We relied on previous studies on interconnects~\cite{litz2008velo,choi2016quantitative} to estimate the delay of
the communication channel.

\citet{litz2008velo} encountered a latency between 36 to 64 cycles to send one packet with 
HyperTransport on a CPU-FPGA platform.\footnote{\citet{litz2008velo} designed an FPGA card with the HTX3 
interface, which is needed for point-to-point communication with HyperTransport. Xilinx used to sell such products but 
they are now discontinued.} Even with a small clock rate, for 
example \SI{500}{\mega\hertz}, we can expect a latency of around 72 to \SI{128}{\nano\second}, close to an uncached 
memory access. \citet{choi2016quantitative} have similar results with QPI-based 
platforms.\footnote{\citet{choi2016quantitative} had access to a QPI-based CPU-FPGA platform thanks to a 
collaboration between Intel and academics at that time.}

Hence, we simulated a delay of \SI{128}{\nano\second} to send one packet.
This corresponds to the worst-case scenario to send one
packet. Since the reference latency we have for AMD HyperTransport and Intel QPI are for FPGA prototypes, lower 
latencies are expected with an ASIC implementation.

Furthermore, since we use a point-to-point connection, we did not consider a fluctuation of the latency. Moreover, as 
explained in~\autoref{subsubsec:implementation:channel:fifo}, only one 
core of the main CPU is running while in SMM\@.

Finally, we simulate the same interconnect and delay between the main CPU and the FIFO, and between
the co-processor and the
FIFO.\footnote{In practice, one would need to use a similar interconnect or a glue logic for the ARM architecture.}

\subsubsection{SMI handlers}
\label{subsubsec:evaluation:setup:smi_handlers}
For our performance evaluation, we used SMI handlers from EDK~II and coreboot.
EDK~II does not implement any hardware initialization nor vendor-related SMI handlers.
At the time of writing, most of SMI handlers available in EDK~II at runtime are dependent on hardware components that 
cannot be easily simulated (\eg an Opal device or a TPM chip).

In our evaluation, we used the VariableSmm SMI handlers from EDK~II\@. They manage
variables within the \acrshort{SMM}~\cite{yaotour} thanks to four different handlers: GetVariable, SetVariable, 
QueryVariableInfo and GetNextVariableName (GNVN).

Since coreboot provides hardware initialization and vendor-related SMI handlers, we use them for our evaluation.
In addition, these handlers communicate with devices, which can be simulated with gem5. A majority of these handlers 
are simpler compared to the VariableSmm SMI handlers.
We used three SMI handlers for the Intel ICH4 i82801gx\footnote{\Eg present on motherboards from Apple,
ASUS, GIGABYTE, Intel or Lenovo.} and two for the AMD Agesa Hudson southbridge.\footnote{\Eg present on
motherboards from AMD, ASUS, HP, Lenovo or MSI.} These SMI handlers process hardware events such as
pressing the power button (PM1), General Purpose Events (GPE), Advanced Power Management Control (APMC) events, or Total 
Cost of Ownership (TCO) events.

\subsection{Security evaluation}
\label{subsec:evaluation:security}
There is no public dataset of vulnerable SMM code, in contrast to userland applications.
Attacks targeting the \acrshort{SMM} are highly specific to the architecture and to the proprietary 
code of the platform. Such code is therefore not publicly available and would not execute on our 
experimental setup, thus cannot be used to test our solution.

Consequently, we have implemented SMI handlers with vulnerabilities similar to previously disclosed ones 
(see~\autoref{subsec:smm}) affecting real-world firmware. We reproduced attacks exploiting the following
vulnerabilities giving arbitrary execution: (1) A buffer overflow in a SMI handler allowing an attacker to modify the 
return address stored on the stack~\cite{kallenberg2013defeating}; (2) An arbitrary write allowing
an attacker to modify a function pointer used in an indirect call~\cite{nucSMMVuln}; (3) An arbitrary write allowing an
attacker to modify the SMBASE~\cite{sogetiSMMVuln}; and (4) An insecure indirect call where the function pointer is
retrieved from a data structure controlled by the attacker~\cite{thinkpwn}.

\begin{table}[ht]
    \centering
    \setlength{\tabcolsep}{3pt}
    \resizebox{\columnwidth}{!}{
    \begin{tabular}{lllc}
        \toprule
        Vulnerability   & Attack Target    & Security Advisories                & Detected \\
        \midrule
        Buffer Overflow & Return address   & CVE-2013-3582~\cite{CVE_2013_3582} & Yes \\
        Arbitrary write & Function pointer & CVE-2016-8103~\cite{CVE_2016_8103} & Yes \\
        Arbitrary write & SMBASE           & LEN-4710~\cite{len_4710}           & Yes \\
        Insecure call   & Function pointer & LEN-8324~\cite{len_8324}           & Yes \\
        \bottomrule
    \end{tabular}
    }
    \caption{Effectiveness of our approach against state-of-the-art attacks}
    \label{tbl:attacks_detected}
\end{table}

As shown in~\autoref{tbl:attacks_detected}, the monitor detected all these attacks as soon as it received and processed 
the messages, since these attacks modify the control-flow of the SMM code (\ie its behavior). We did not encounter
false positives, which is expected since we use a conservative strategy regarding indirect calls.
Also, while bad software engineering practices using function type cast could introduce false positives, we did not 
encounter such case in the code we evaluated, as no function cast was present.

While our implementation detects these intrusions, an attacker could theoretically bypass our solution.
First, by managing to send multiple forged
packets without any other legitimate packets being sent in the middle.
Second, doing so without redirecting the control-flow to send these forged packets (an attack out of our threat model,
see~\autoref{sec:threat_model}).

Finally, our CFI implementation performs a sound analysis to recover the potential targets
of an indirect call. Therefore, the analysis is not complete and it would be possible for an attacker 
to redirect the control flow to a function that should have never been called, but that has the 
expected type signature.
Nonetheless, we argue that a type-based \acrshort{CFI} increases the difficulty for the attacker, 
since the only available targets for an indirect call are a subset of the existing functions within 
the SMRAM with the right type signature. Our analysis with EDK~II gave 158 equivalence 
classes of size 1, 24 of size 2, 42 of size 3, 2 of size 5, 1 of size 9, and 1 of size 13. As mentioned
by \citet{burow2017control}, a high number of small equivalence classes provides
a precise CFG\@.
A way to improve the precision of the CFG would be to combine our static analysis (providing some context-sensitivity 
with the type information), with a points-to analysis, such as the work from \citet{DSA}.
Such points-to analysis can sometimes give the complete set of the functions being called at an indirect call site.
An idea would be that if the points-to analysis gives a complete set, the monitor uses this information to validate
an indirect call, otherwise it uses the over-approximation of the type signature.

\subsection{Performance evaluation}
\label{subsec:evaluation:performance}
As explained in~\autoref{subsubsec:implementation:channel:existing}, the time spent in SMM is limited (threshold
of \SI{150}{\micro\second})~\cite{intelBITS150,delgado2013performance}.
On that account, we evaluated the running time overhead of our solution on SMI handlers 
for the main CPU\@. We also evaluated the time it takes for the co-processor to process the messages sent by different 
SMI handlers. Thus, we can estimate the time between an intrusion, its detection, and remediation.

Finally, the size of firmware code is limited by the amount of flash (\eg 8MB or 16MB). Thus, we 
evaluated the size of the firmware before and after our instrumentation.

\subsubsection{Runtime overhead}
The additional SMM code added with our instrumentation introduces two costs: the raw communication delay between the
main CPU and the hardware FIFO; and the instrumentation overhead. The former is related to the time it takes the main 
CPU to push the packets to the FIFO\@. The latter is due to multiple factors, such as fetching and executing new 
instructions or storing intermediate values resulting in register spilling (\eg the return address of a function fetched 
from the stack).

\begin{figure*}[ht]
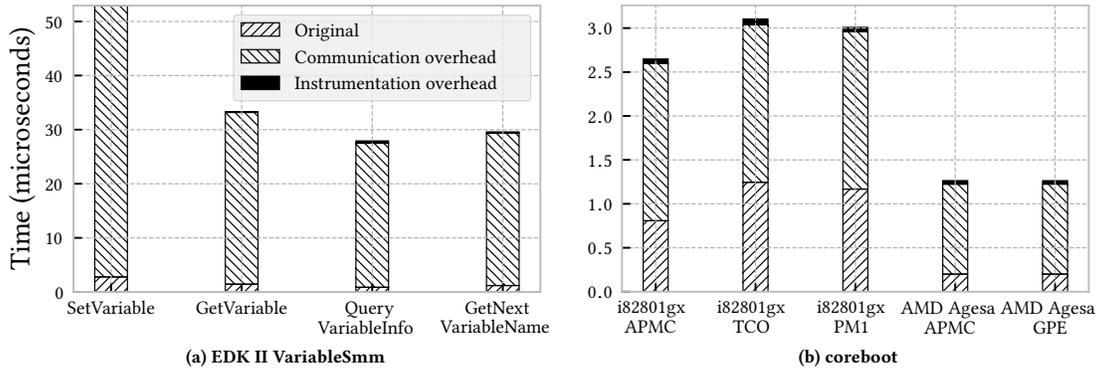

    \centering
    \subfloat[EDK~II VariableSmm]{
        \scalebox{0.9}{
            \input{pgf/perf_proc_edk2.tex}
        }
    }%
    \subfloat[coreboot]{
        \scalebox{0.9}{
            \input{pgf/perf_proc_coreboot.tex}
        }
    }
    \caption{Time (in microseconds) to execute each \acrshort{SMI} handler (averaged over 100 executions) with the 
        original time, and the overhead divided between the communication overhead due to pushing packets to the FIFO 
        and the instrumentation overhead.}
    \label{fig:perf_main_cpu}
\end{figure*}

We performed 100 executions of each SMI handler we selected for our evaluation 
(see~\autoref{subsubsec:evaluation:setup:smi_handlers}). For each SMI handler, we measured the time it takes for the 
original handler to execute, the cost of the communication, and the additional instrumentation overhead.
The results we obtained are illustrated by~\autoref{fig:perf_main_cpu}.

We see that even with a low latency of \SI{128}{\nano\second} there
is a high overhead. It is due to the number of messages related to the shadow stack
(see~\autoref{tbl:count_message_types}), while the number of messages for indirect calls or
the integrity of the relevant CPU registers (SMBASE and CR3) are negligible.
However, this overhead is below the \SI{150}{\micro\second}
threshold~\cite{intelBITS150} ensuring that the impact
on the performance of the system is low and not noticeable for the user.

\begin{table}[ht]
    \centering
    \setlength{\tabcolsep}{3pt}
    \resizebox{\columnwidth}{!}{
        \begin{tabular}{llcccc}
            \toprule
            & \multicolumn{5}{c}{\hspace*{35pt}Number of packets sent} \\ \cmidrule{3-5}
            \multicolumn{2}{l}{\raisebox{-1em}{SMI Handler}} & \specialcell{Shadow\\ stack\\(SS)} & \specialcell{Indirect\\call\\(IC)} & \specialcell{SMBASE\\\& CR3\\(SC)} & \specialcell{Total\\number of\\packets} \\
            \midrule
            \multicolumn{2}{l}{\textbf{EDK~II}} & & & & \\
            \multicolumn{2}{l}{VariableSmm} & & & & \\
            & SetVariable           & 384       & 4       & 4       & 392 \\
            & GetVariable           & 240       & 4       & 4       & 248 \\
            & QueryVariableInfo     & 299       & 4       & 4       & 208 \\
            & GetNextVariableName   & 212       & 4       & 4       & 220 \\
            \midrule
            \multicolumn{2}{l}{\textbf{coreboot}} & & & & \\
            \multicolumn{2}{l}{Intel i82801gx} & & & & \\
            & APMC/TCO/PM1          & 8         & 2       & 4       & 14 \\
            \multicolumn{2}{l}{AMD Agesa Hudson} & & & & \\
            & APMC/GPE              & 4         & 0       & 4       & 8 \\
            \bottomrule
        \end{tabular}
    }
    \caption{Number of packets sent during one \acrshort{SMI} handler (Number of packets per message type: SS=2, IC=2, 
        SC=4)}
    \label{tbl:count_message_types}
\end{table}

\subsubsection{Co-processor performance}
\begin{figure}[ht]
    \resizebox{0.9\columnwidth}{!}{
    \input{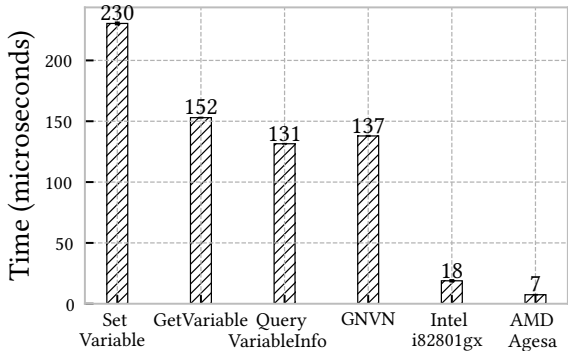}
    }
    \caption{Time (in microseconds) to process all the messages sent by one execution of each SMI handler for the co-processor}
    \label{fig:perf_coproc}
\end{figure}

We measured the time it takes for the monitor to process all the messages generated by one execution of each SMI 
handler. We made an average of 1000 executions. Results are illustrated in~\autoref{fig:perf_coproc}.

For each SMI handler there is at least a factor of 4 between the time it takes for the target to execute the 
instrumented
SMI handler and the time it takes for the co-processor to process all the messages that have been sent by the 
instrumented SMI handler.
For example, we see in~\autoref{fig:perf_main_cpu} that it takes around \SI{52}{\micro\second} 
to execute the SetVariable SMI handler, and in~\autoref{fig:perf_coproc}  that it takes around 
\SI{230}{\micro\second} to process all the messages. This means that there is a delay 
between an intrusion and its detection, but such delay will be less than a millisecond.
Hence, the co-processor could start a remediation action within one millisecond after an intrusion occurred.

In our threat model, the attacker already has kernel privileges before attacking the SMM code. Our objective is not 
to detect intrusions that could have been done solely with kernel privileges, such as leaking
confidential data. We consider that the final objective of the attacker is to remain persistent in the system even in 
the case of a reboot. In this case, the remediation action does not have to be taken immediately. This remediation would 
not prevent the intrusion, but will recover to a safe state.

\subsubsection{Firmware size}
For EDK~II, our instrumentation added $17 408$ bytes to the firmware code. However, firmware is compressed 
before being stored in the flash and only a subset of the firmware is related to the SMM\@. We measured a
0.6\% increase in size of the compressed firmware.
Thus, our instrumentation incurs an acceptable overhead in terms of size for the firmware.

For coreboot, our instrumentation added $568$ bytes for the AMD Agesa Hudson SMI handlers and $3 448$ bytes for 
the Intel i82801gx SMI handlers. However, we were not able to measure 
the whole firmware size when building coreboot with our LLVM toolchain, since coreboot does not support clang as a
compiler.\footnote{We could measure the size of coreboot compiled with gcc, but the size varies when using clang or 
gcc.} We built separately the SMI handlers from coreboot toolchain for our evaluation, but compiling the whole 
firmware (not just the SMM related code) is not possible.

\section{Related work}
    \label{sec:related_work}
    \newcommand{\tikzcircle}[2][black,fill=black]{\tikz[baseline=-0.5ex]\draw[#1,radius=#2] (0,0) circle ;}%
\newcommand{\xmark}{\tikzcircle[black, fill=white]{2.5pt}}
\newcommand{\cmark}{\tikzcircle{2.5pt}}
\begin{table}[ht]
    \centering
    \begin{tabular}{llllllll}
        \toprule
                 & \multicolumn{7}{c}{Approach}\\ \cmidrule{2-8}
        \raisebox{-1cm}{Property}
        & \rotatebox[origin=c]{90}{Our approach}
        & \rotatebox[origin=c]{90}{Copilot~\cite{2004copilot}}
        & \rotatebox[origin=c]{90}{DeepWatch~\cite{bulygin2008chipset}}
        & \rotatebox[origin=c]{90}{HyperSentry~\cite{azab2010hypersentry}}
        & \rotatebox[origin=c]{90}{KI-Mon~\cite{lee2013ki}}
        & \rotatebox[origin=c]{90}{MGuard~\cite{mguard}}
        & \rotatebox[origin=c]{90}{Hardware CFI~\cite{intelCET,lee2015towards}} \\
        \midrule
        Can monitor SMM code        & \cmark & \xmark & \cmark & \xmark & \cmark & \cmark & \cmark \\
        Flexible                    & \cmark & \cmark & \cmark & \cmark & \cmark & \cmark & \xmark \\
        No semantic gap issue       & \cmark & \xmark & \xmark & \cmark & \xmark & \xmark & \cmark \\
        Detect transient attacks    & \cmark & \xmark & \xmark & \xmark & \cmark & \cmark & \cmark \\
        No new/modified hardware    & \xmark & \xmark & \cmark & \cmark & \xmark & \xmark & \xmark \\
        \bottomrule
    \end{tabular}
    \caption{Comparison between our work and the related work. \cmark~: has the property \xmark~: does not have the property}
    \label{tbl:related_work_comparison}
\end{table}

In this section, we discuss existing approaches to monitor low-level components such as firmware and 
kernels, using a hardware-based approach.
To the best of our knowledge, the only commercially available technology that offers SMM integrity 
monitoring is HP Sure Start~\cite{hp2017ss3amd,hp2017ss3intel}.
It uses the chipset and the CPU to monitor SMM code integrity and relies on additional hardware to
take actions per a predefined policy. The details of its implementation, however, are not public. 
Thus, we cannot compare it in details to other approaches in the literature.

The other approaches presented are not necessarily focused on monitoring the 
SMM (\ie the firmware), but they could be adapted to that aim.
We can distinguish two different types of approaches: snapshot-based approaches are presented 
in~\autoref{subsec:related:snapshot} and event-based approaches in~\autoref{subsec:related:event}.
We summarize the comparison between these approaches and our work
in~\autoref{tbl:related_work_comparison}.

\subsection{Snapshot-based approach}
\label{subsec:related:snapshot}
The first approach consists in taking periodic snapshots of all or any part of the target state and then to analyze 
these snapshots to detect intrusions.

To the best of our knowledge, \citet{zhang2002secure} were the first to propose a co-processor for
intrusion detection using a snapshot-based approach. However, they did not implement their design.

Notable implementations of such approach are Copilot~\cite{2004copilot}, DeepWatch~\cite{bulygin2008chipset},
and HyperSentry~\cite{azab2010hypersentry}. Copilot is a kernel integrity monitor using a co-processor on a 
\acrshort{PCI} card to take periodic snapshots of the main memory. The authors also described how to 
write rules describing the relationships between kernel objects to detect the presence of kernel 
rootkits~\cite{petroni2006architecture}. Copilot, however, cannot monitor the SMM since it does
not have access to SMRAM\@.
DeepWatch uses a similar approach to Copilot, but the monitor runs on an embedded core in the 
chipset, which allows the monitoring of the SMM\@.
HyperSentry leverages the \acrshort{SMM} to perform measurements giving access to the CPU-context, 
but it impedes its ability to monitor the SMM itself.

In general, these snapshot-based solutions are unable to detect 
transient attacks, where an attacker does not make persistent changes 
(\eg one could erase its traces before each snapshot).

\subsection{Event-driven approach}
\label{subsec:related:event}
All the following event-driven approaches, like our approach, require a new specific hardware component or a 
modification of an existing hardware component.

Vigilare~\cite{vigilare} snoops the memory bus traffic of 
the host by using an external hardware component to detect modifications of immutable regions of a kernel. This approach
does not suffer from transient attacks: as soon as an illegal modification is made it is detected.
KI-Mon~\cite{lee2013ki}, its successor, also monitors mutable kernel objects.
MGuard~\cite{mguard} follows a similar approach, but incorporates the integrity
monitor inside a DRAM DIMM device. One limitation,
however, that affects these solutions, is their inability to access the CPU state of the host they are monitoring
(semantic gap issue). \citet{jang2014atra} demonstrated the practicability of an evasion scheme by
modifying the CR3 register. Our solution, while being event-driven, is not vulnerable to the
CR3 attack since we monitor it as explained in~\autoref{subsubsec:implem:instru:invariants}.

While our generic approach does not focus on CFI, our evaluation used CFI as a detection method to demonstrate the
applicability of our approach. Thus, we compare our solution with hardware-based CFI approaches.

\citet{lee2015towards} use a co-processor and debugging
features available in ARM processors to enforce \acrshort{CFI} on the main CPU\@. \citet{davi2015hafix}
extend the instruction set of the processor to enforce a similar policy. A recent document from
Intel~\cite{intelCET} suggests that future Intel processors will have a backward-compatible \acrshort{CFI} technology in
hardware (and available for the \acrshort{SMM}).

Hardware-based CFI approaches modify the processor or use additional hardware solely to
enforce CFI\@. Our approach, on the other hand, is more flexible since different detection 
approaches could be implemented without modifying our hardware component. The flexibility of the solution is important, 
because the types of vulnerabilities exploited evolves over time.

As discussed in~\autoref{sec:background}, CFI can be implemented without hardware modifications by inlining the 
detection logic inside the target. Our approach, however, isolates the critical parts of the detection process 
such as the shadow call stack and the indirect call mappings.

\section{Conclusion and future work}
    \label{sec:conclusion}
    In this work, we propose a new event-based approach for low-level software using three key
components: a co-processor to isolate the monitor, a communication channel to 
reduce the semantic gap, and an instrumentation of the software to enforce the 
communication.
We show that this approach can be followed to detect intrusions targeting SMM code by using CFI and by
ensuring the integrity of relevant CPU registers (CR3 and SMBASE).
However, it can implement different detection methods.
Unlike other approaches, we solve the challenges of the semantic gap and the transient attacks while
remaining flexible.

We implemented our approach by instrumenting and monitoring real-world firmware.
The results show that we detect state-of-the-art attacks against the \acrshort{SMM}, while remaining below the
\SI{150}{\micro\second} threshold, thus avoiding any noticeable impact on the user.

For future work, we would like to investigate how we could leverage such a co-processor-based
monitor to (1) start remediation strategies and study the impact on the user experience; and (2) apply our approach to 
monitor other targets and use other detection methods.
For example, our approach could be used to monitor ARM TrustZone secure world~\cite{ARMTZ}, since it 
offers a similar environment than SMM (\eg a non-secure bit to know whether the CPU is in the secure world or not, like 
the SMIACT\# signal).

\appendix
\section{gem5}
    \label{appendix:gem5}
    \begin{table}[ht]
    \centering
    \begin{tabular}{llll}
        \toprule
        \multicolumn{2}{l}{Parameter}               & x86                    & ARM Cortex A5 \\
        \midrule
        \multicolumn{2}{l}{CPU Type}                & DerivO3Cpu             & timing\parnote{We use the timing model since the A5 is a single-issue in-order CPU and our evaluation mainly depends on load/store operations.} \\
        \multicolumn{2}{l}{Clock}                   & \SI{2}{\giga\hertz}    & \SI{500}{\mega\hertz} \\
        \multicolumn{2}{l}{Restricted FIFO latency} & \SI{128}{\nano\second} & \SI{128}{\nano\second} \\
        \multicolumn{2}{l}{Cache line size}         & 32 B                   & 32 B    \\\cmidrule{2-4}
        \multirow{2}{*}{L1 I}& Size                 & 32 KB                  & 16 KB\parnote{\label{ftn:cachesize} The cache size has a range of options: 4 KB, 8 KB, 16 KB, 32 KB or 64 KB.} \\
                             & Associativity        & 2                      & 2 \\\cmidrule{2-4}
        \multirow{2}{*}{L1 D}& Size                 & 64 KB                  & 16 KB\textsuperscript{\ref{ftn:cachesize}} \\
                             & Associativity        & 2                      & 4 \\\cmidrule{2-4}
        \multirow{2}{*}{L2}  & Size                 & 2 MB                   & 512 KB \\
                             & Associativity        & 8                      & 8 \\\cmidrule{2-4}
        \multirow{2}{*}{DRAM}& Type                 & DDR3\_1600             & LPDDR3\_1600\_x32\parnote{Educated guess, based on the fact that this is a standard for low power consumption memory.} \\
                             & Size                 & 1024 MB                & 10 MB \\
        \bottomrule
    \end{tabular}
    \parnotes
    \caption{Parameters used with gem5 for the x86 and the ARM simulation}
    \label{tbl:gem5_config}
\end{table}

In~\autoref{tbl:gem5_config}, we show the different parameters used to configure gem5 for the simulation of the main CPU 
and the co-processor. We use the default parameters of the out-of-order x86 simulation, except the CPU clock, which we 
set to a higher frequency. For the ARM Cortex simulation, we derived the parameters from the ARM technical reference 
manual~\cite{ARMCortexA5}.

\begin{acks}
    The authors would like to thank and acknowledge the contribution of the following people (in
    alphabetical order) for their helpful comments, technical discussions, feedback and proofing of 
    earlier versions of this paper:
    Vali Ali,
    Boris Balacheff,
    Pierre Belgarric,
    Rick Bramley,
    Chris Dalton,
    Carey Huscroft and
    Jeff Jeansonne.
    In addition, we would like to thank the anonymous reviewers for their feedback.
\end{acks}

\bibliographystyle{ACM-Reference-Format}
\bibliography{paper}

\end{document}